\documentclass{appolb}
\usepackage{epsfig}
\usepackage{cite}
\usepackage{color}

\begin{document}
\title{ Domain structure created by irreversible adsorption of dimers
\thanks{Presented at the XXV Marian Smoluchowski Symposium on Statistical Physics,
Kraków, Poland, September, 2012.}
}
\author{ Jakub Barbasz$^{1, 2}$ \thanks{e-mail: ncbarbas@cyf-kr.edu.pl} and Micha\l{} Cie\'sla$^{2}$ \thanks{e-mail: michal.ciesla@uj.edu.pl}
\address{$^1$Institute of Catalysis and Surface Chemistry, Polish Academy of Sciences, 30-239 Kraków, Niezapominajek 8, Poland.}
\address{$^2$M. Smoluchowski Institute of Physics, Jagiellonian University, 30-059 Kraków, Reymonta 4, Poland.}
}
\maketitle
\begin{abstract}
Structure of monolayers built during adsorption process is strongly related to the properties of adsorbed particles. The most important factor is their shape. For example, adsorption of elongated molecules on patterned surfaces may produce certain orientational order inside a covering layer. This study, however, focuses on random adsorption of dimers on flat, homogeneous surfaces. It has been observed that despite the lack of global orientational ordering, adsorbed dimers may form local, orientationally ordered structures \cite{bib:Ciesla2012a, bib:Ciesla2013b}. Our investigations focus on the dependence between domain size distribution and environmental parameters such as ionic strength, which affects the range of electrostatic interaction between molecules.
\end{abstract}
\PACS{05.45.Df, 68.43.Fg}
\section{Introduction}
Irreversible adsorption of particles at solid and liquid interfaces is of major significance for many fields such as medicine and material sciences as well as pharmaceutical and cosmetic industries e.g. \cite{bib:Rafati2012, bib:Galli2001}. For example adsorption of some proteins plays crucial role in blood coagulation, fouling of contact lenses, and plaque formation. The process is also used in membrane filtration units.  Additionally, controlled adsorption is fundamental for efficient chromatographic separation and purification, gel electrophoresis as well as biosensing and immunological assay performance.
\par
Adsorption layers formed on flat, homogeneous surface typically exhibit random structure without any signs of order. However, a number of latest studies on adsorption focuses on deposition on specially prepared, patterned surfaces \cite{bib:Browne2004, bib:Bae2005, bib:Cho2012, bib:Keller2011, bib:Zemla2012}. It is due to the possibility of gaining better control over adsorption process and the properties of resulting mono- and multilayers. In such experiments, adsorbed particles take on order of the base. On the other hand, anisotropic particles can form orientationally ordered layers even on flat, homogeneous surfaces. Ordering can originate from non-uniform alignment of particles in a bulk phase \cite{bib:Ciesla2013a} or appear spontaneously in a form of ordered domains. The second is the subject of investigation in our study.  As far as we know, such ordering hasn’t been investigated before.  Although local ordering has been reported in earlier numerical studies on dimer adsorption \cite{bib:Ciesla2012a, bib:Ciesla2013b}, not even basic properties of orientationally ordered domains have been further investigated. Here, we focus on domain sizes and their spatial distribution for different range of repulsive interaction between adsorbed particles.
\section{Model}
\begin{figure}[htb]
\vspace{1cm}
\centerline{%
\includegraphics[width=2cm]{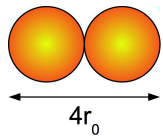}}
\caption{Model of dimer particle used in Random Sequential Adsorption simulations.}
\label{fig:model}
\end{figure}

Adsorbed particle is modeled by two identical touching spheres of radius $r_0$ (see Fig.\ref{fig:model}). 
This model was chosen because it is the simlest numerical model of anisotropic particle. Moreover, it has been carefully studied and basic properties of the adsorbed layers are well known \cite{bib:Ciesla2012a, bib:Ciesla2013b}.  It can be also used as a coarse-grained model for some biomolecules \cite{bib:Tozzini2005}. Electrostatic interaction between two dimers originates from electrostatic potentials generated by individual spheres. In typical substrate this interaction is approximated by an exponentially decaying Yukawa potential known also as screened electrostatic potential \cite{bib:Senger2000, bib:Trulsson2009}:
\begin{equation}
\label{uel}
U_{el}(r) = \left\{
 \begin{array}{cc} 
    - \frac{Q_0 q^2 r_0^2}{4\epsilon_0 \epsilon_r r}  \exp\left[-\frac{r - r_0}{L_e}\right] & \mbox{for $L_e >0$}, \\
   0 & \mbox{for $L_e = 0$},
  \end{array}
  \right.
\end{equation}
where $a$ and $q$ are particle diameter and charge, respectively; and $r$ is distance between individual spheres from different particles. $Q_0$ is a constant parameter, $\epsilon_0$ is a vacuum permittivity and $\epsilon_r$ characterizes dielectric properties of a solvent. Range of the interaction is controlled by parameter $L_e$, known also as the Debye screening length. It is a key parameter for electrostatic interaction in electrolytes and can be calculated by \cite{bib:Debye1923, bib:Adamczyk-book}:
\begin{equation}
Le=\sqrt{\frac{\epsilon k_B T}{2e^2 I}},
\end{equation}
where $k_B$ is the Boltzmann constant, $T$ is temperature, $e$ is elementary charge, and $I$ is ionic strength of electrolyte solution. 
\par
Adsorption layers were generated using Random Sequential Adsorption algorithm introduced by Feder \cite{bib:Feder1980}. It is based on repeating the following steps:
\begin{description}
\item[I] a new virtual dimer is created on a collector. Its position and orientation are selected randomly according to uniform probability distribution;
\item[II.a] the virtual particle undergoes overlapping test with its nearest neighbors;
\item[II.b] the total electrostatic potential $U$ between the virtual molecule and previously adsorbed dimers is calculated using Eq.(\ref{uel});
\item[III] if there is no overlap the virtual particle is added to the existing layer with a probability $P(U) = \exp(−U/k_B T)$ \cite{bib:Adamczyk1990, bib:Oberholzer1997}. Otherwise the virtual dimer is removed and abandoned. 
\end{description}
The whole procedure is repeated for a specified number of times, which guarantees close approach to the jamming limit. The simulation of adsorption process was performed for a square collector of $200a$-side size and was stopped after $6.37\cdot 10^8$ iterations. For each set of parameters, 20 to 100 independent simulations were performed. Parameters of electrostatic potential (\ref{uel}) were chosen based on \cite{bib:Ciesla2013b} to describe typical experimental conditions of water solutions. Therefore, the thermal energy $k_BT$ at a room temperature acts as energy unit and $\epsilon_r=78$. The $r_0=9.29 \,\, nm$ provides length scale typical to mid sized bio-molecules, and $Q_0 = 6.78 \,\, k_B T a / e^2$. 
\subsection{Definition of domain}
\begin{figure}[htb]
\vspace{1cm}
\centerline{%
\includegraphics[width=8cm]{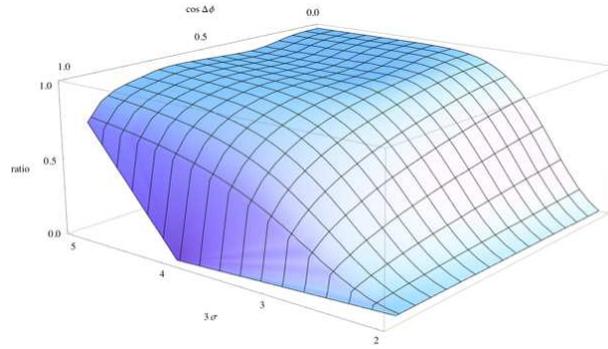}
}
\caption{Ratio of particles in domains as a function of distance and angle difference used in cluster classification procedure. $L_e = 1.0$.}
\label{fig:ratio}
\end{figure}

Particle is treated as a domain member when it is close to other  domain components and its orientation does not differ much from the mean domain orientation. However, specific threshold values of distance and angle may have significant impact on domain size and shape. Moreover, mean distance between particles in a layer depends on electrostatic repulsion range $L_e$. It means that the same distance threshold value for all $L_e$ will prevent domains to be classified as such when interaction range is large. On the other hand, such "soft" interaction affects the effective anisotropy of a particle which changes the tendency for domain formation. Fig.\ref{fig:ratio} shows ratio of particles classified as domain members in respect to specific values of distance and angle difference used in classification process.
\par
For the purpose of the present study, the concept of domain is defined according to Refs.\cite{bib:Fujiwara1998, bib:Esselink1990}. Particle is domain member when it fulfills both of the following conditions:
\begin{description}
\item[I] its angle does not differ from mean domain orientation more than $15^{\circ}$;
\item[II] its monomer-to-monomer distance to the closest domain component is not larger than $3.0 \sigma$, where $\sigma$ is characteristic size of an interacting dimer.  Distance is measured between the centres of monomers. 
\end{description}
Parameter $\sigma$ depends on $L_e$. For hard core interaction ($L_e=0$), $\sigma=r_0$. With the growth of $L_e$ parameter $\sigma$ increases, to compensate lower coverage ratio $\theta$ of a layer:
\begin{equation}
\sigma(L_e) = \sqrt{\frac{\theta(L_e=0)}{\theta(L_e)}}\cdot r_0.
\label{eq:sigma}
\end{equation}
For convenience, $\sigma$ values used in this study is collected in Tab.\ref{tab:sigma}.
\begin{table}[htb]
  \begin{center}
  \begin{tabular}{ccc}
  $L_e$ & $\theta$ [$r_0$] & $\sigma$ [$r_0$] \\
  \hline
  $0.0$	& $0.547$	& $1.000$	\\
  $0.1$	& $0.510$	& $1.036$	\\
  $0.2$	& $0.485$	& $1.062$	\\
  $0.5$	& $0.435$	& $1.121$	\\
  $1.0$	& $0.372$	& $1.213$	\\
  $2.0$	& $0.277$	& $1.405$	\\
  $5.0$	& $0.146$	& $1.936$	\\
  \hline
  \end{tabular}
  \caption{Parameter $\sigma$ (\ref{eq:sigma}) for different electrostatic interaction range $L_e$. Values of maximal random coverage ratios $\theta$ were taken from \cite{bib:Ciesla2013b}.}
  \end{center}
  \label{tab:sigma}
\end{table}
Using the above definition of a cluster, typically, $25\%$ to $45\%$ of dimers are classified as domain members. 
\section{Results and discussion}
\subsection{Hard core interaction}
\begin{figure}[htb]
\vspace{1cm}
\centerline{%
\includegraphics[width=4cm]{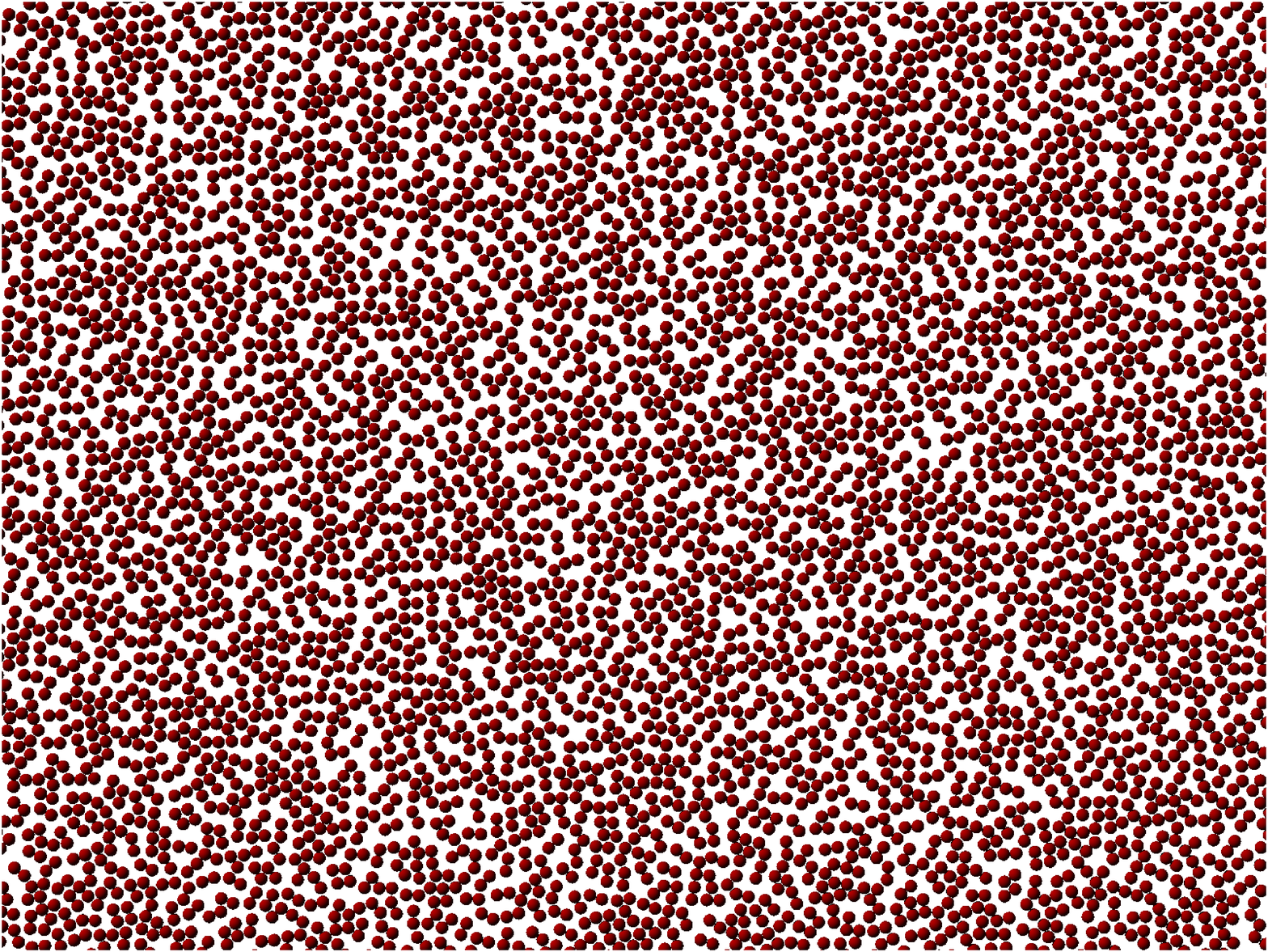}
\includegraphics[width=4cm]{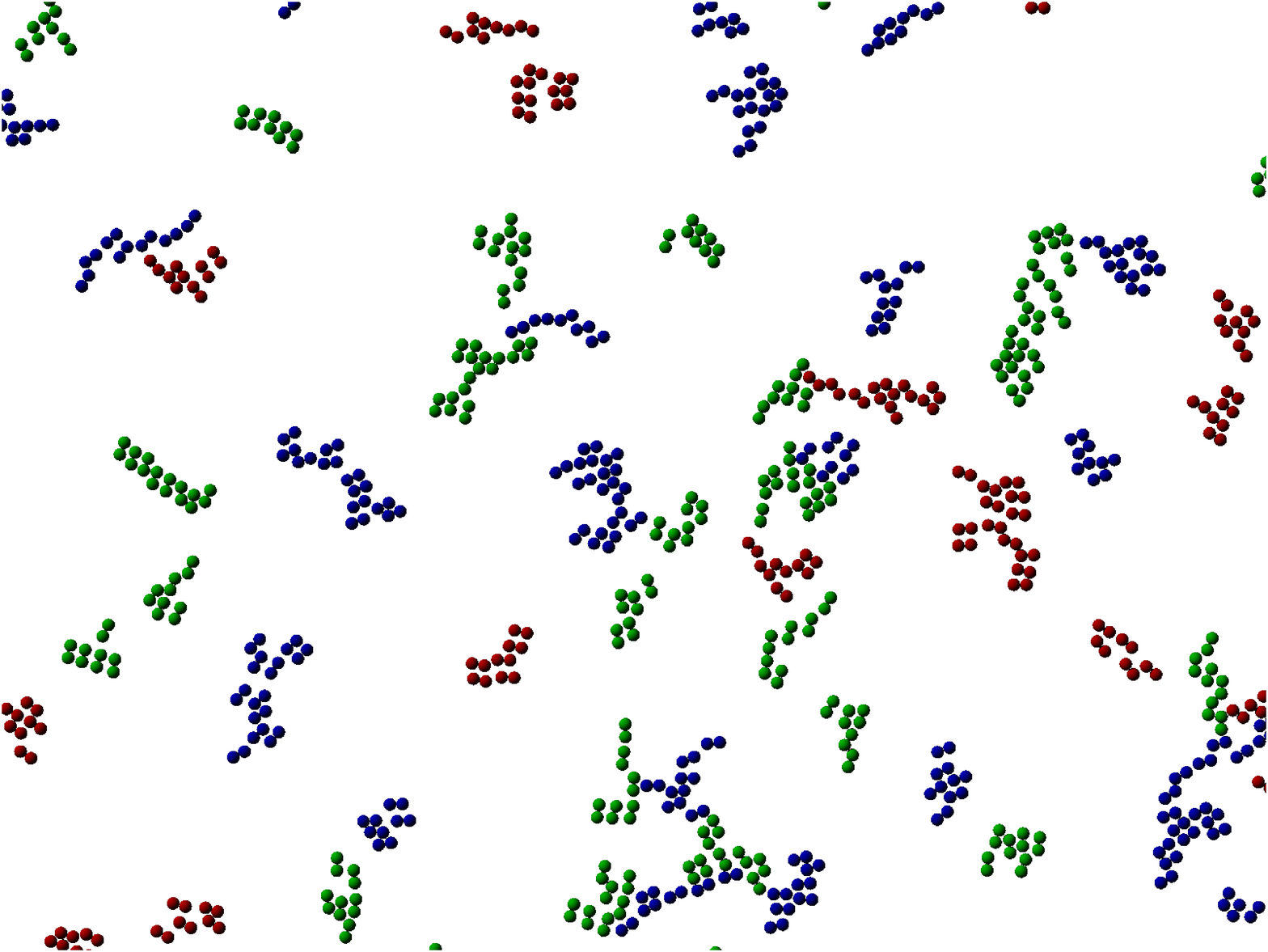}
}
\caption{Fragment of layer build of adsorbed dimers (left) and the same fragment but only clusters including at least 5 particles are shown (right). Cluster colour describes its mean orientation. $L_e = 0$.}
\label{fig:clusters}
\end{figure}

\begin{figure}[htb]
\vspace{1cm}
\centerline{%
\includegraphics[width=6cm]{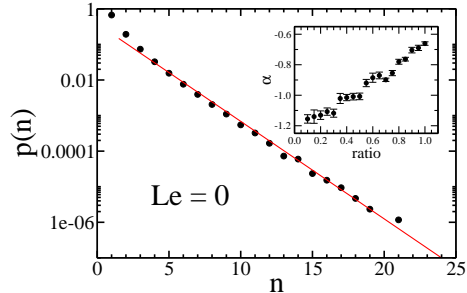}
}
\caption{Histogram of domain sizes in hard core ($L_e = 0$) interacting dimer layers. Dots are measured data whereas solid line is an exponential fit: p(n) = $0.379\cdot \exp (-0.631 \cdot n)$. Insets shows fitted exponentials to domain size distribution for partially filled layers.}
\label{fig:hist00}
\end{figure}

Adsorption of dimers not interacting electrostatically ($L_e=0$) provides reference level for further investigations. The first important observation is that appearance of orientationally ordered dimers clusters is typical for adsorbed layers generated by RSA algorithm (see Fig.\ref{fig:clusters}).
The distribution of domain sizes is presented in Fig.\ref{fig:hist00}.
 Its exponential decay suggests that there is constant probability of classifying particle as a member of orientationally ordered cluster. The probability read from plot is equal to $0.532 \pm 0.005$. This value is much higher than $0.167$, which is the probability of random selection of particle orientation within $[-15^{\circ}, 15^{\circ}]$ interval. On the other hand, a single cluster borders on more than one dimer, which increases the probability of finding nearby parallely aligned particle. Therefore, the above results do not answer the question whether domain formation is a merely statistical effect or rather intrinsic property of elongated particle adsorption process. To check it, we analysed cluster statistic before the maximal random coverage has been reached. For result comparison purposes, domain classification criteria had to be changed due to lower particle density according to (\ref{eq:sigma}). For example, $\sigma=\sqrt{2}$ for half of particles inside a layer. Results presented in Fig.\ref{fig:hist00} inset show that denser layers are characterized by larger probability of parallel alignment. For example, after adsorption of only $10\%$ of the particles, the probability of classifying particle as a domain component is $0.317 \pm 0.006$, which is significantly smaller than the value for maximally covered layers. Therefore, it can be concluded that Random Sequential Adsorption of anisotropic particles leads to formation of orientationally ordered clusters even in environment with no specific direction.
\subsection{Screened electrostatic interaction}
\begin{figure}[htb]
\vspace{1cm}
\centerline{%
\includegraphics[width=6cm]{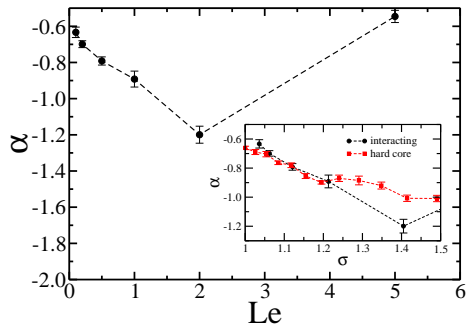}
}
\caption{Fitted exponentials to domain size distribution for electrostatically repulsing dimers. Insets shows the same exponentials dependence on parameter $\sigma$ for both: hard core and electrostatically interacting dimers.}
\label{fig:a_le}
\end{figure}

Anisotropic molecules characterised by soft interactions also form orientationally ordered clusters. Moreover, the domain size distribution here is exponential, too.  The probability of attaching particle to cluster decreases with interaction range; however, it has incomprehensible high value for $L_e=5.0$, which requires further study (see Fig.\ref{fig:a_le}).
Besides the one result, obtained exponents are similar to the ones for hard-core interacting dimers at earlier stage of adsorption ($\sigma>1.0$). It suggests that the main factor affecting formation of orientationally ordered clusters is the density of particles inside layers. The result is not as entirely expected, as soft interaction generated by screened electrostatic potential effectively decreases anisotropy of dimers, as reported in \cite{bib:Ciesla2013b}. On the other hand, soft interaction increases distance, at which shape anisotropy can be noticed by other particles. As shown, both effects compensate each other, at least for $L_e<1$.
\section{Conclusions}
Dimer adsorption on homogeneous flat surfaces was investigated numerically using Random Sequential Adsorption (RSA) algorithm. Formed layers contain orientationally ordered clusters of particles. Their size distribution is exponential. The exponent depends mainly on particle density inside a layer but not on the character of interaction between particles, at least when interaction range is small.
\subsection*{Acknowledgments}
This work was supported by grant MNiSW/0013/H03/2010/70.

\end{document}